\documentclass[a4paper,reqno,12pt,final]{amsart}
\usepackage{amsmath,amsfonts,amssymb,amsthm,amstext,eucal}
\usepackage{graphicx,lscape}
\usepackage[latin1]{inputenc}
\renewcommand{\d}{\partial}
\newcommand{\fl}{\mathfrak{l}}
\newcommand{\fm}{\mathfrak{m}}
\newcommand{\fp}{\mathfrak{p}}
\newcommand{\ft}{\mathfrak{t}}
\newcommand{\fspin}{\mathfrak{spin}}
\newcommand{\fso}{\mathfrak{so}}
\newcommand{\fsu}{\mathfrak{su}}
\newcommand{\fsp}{\mathfrak{sp}}
\newcommand{\eM}{\mathcal{M}}
\newcommand{\EE}{\mathbb{E}}
\newcommand{\half}{\tfrac{1}{2}}

\newcommand{\RR}{\mathbb{R}}

\newcommand{\ZZ}{\mathbb{Z}}

\DeclareMathOperator{\dvol}{dvol}
\begin{document}
\title[supersymmetric fluxbranes]%
{Generalised supersymmetric fluxbranes}
\author[Figueroa-O'Farrill]{José Figueroa-O'Farrill}
\address{Department of Mathematics and Statistics, The University of
  Edinburgh, Scotland, United Kingdom}
\email{j.m.figueroa@ed.ac.uk}
\author[Simón]{Joan Simón}
\address{The Weizmann Institute of Physical Sciences, Department of
  Particle Physics, Rehovot, Israel}
\email{jsimon@weizmann.ac.il}
\thanks{EMPG-01-17, WIS/20/01-OCT-DPP}
%\date{\today}
\begin{abstract}
  We classify generalised supersymmetric fluxbranes in type II string
  theory obtained as Kaluza--Klein reductions of the Minkowski space
  vacuum of eleven-dimensional supergravity.  We obtain two families
  of smooth solutions which contains all the known solutions, new
  solutions called nullbranes, and solutions interpolating between
  them.  We explicitly construct all the solutions and we study the
  U-duality orbits of some of these backgrounds.
\end{abstract}

\maketitle

\tableofcontents

\section{Introduction and conclusions}

There has recently been a lot of interest in the embedding of the
Melvin universe \cite{Melvin} in M/string theory \cite{GibbonsMaeda},
in particular in the type IIA flux 7-brane (F7-brane), whose M-theory
description \cite{DGGH1,DGGH2,DGGH3,GSflux} strongly suggests that
type IIA string theory with magnetic field $M_s^2/g_s^2$ is dual to
type 0A string theory \cite{CG,BGType0}.  They also play an important
role in the supergravity description of the expansion of a D$p$-brane
into a spherical D($p+2$)-brane due to the dielectric effect
\cite{CHC} and in the possible stabilisation of tubular branes
\cite{EmparanFlux,BrecherSaffin}.  The study of analogous magnetic
backgrounds allowing a conformal field theory description
\cite{RT1,RT,TseytlinFlux} provides a framework to study the decay of
unstable backgrounds into stable supersymmetric ones since they are
smoothly connected to the supersymmetric closed string vacuum.
Related work can also be found in
\cite{MotlMelvin,Saffin,CGS,SY,Suyama,TU}.

It was pointed out in \cite{GSflux,Uranga} that when several magnetic
parameters are taken into account the corresponding supergravity
background may preserve some amount of supersymmetry.  One of the
goals of this paper is to classify these possibilities, or
equivalently, to classify supersymmetric fluxbranes in string theory.
To do so, we shall reexamine the geometrical setting giving rise to
the flux-fivebrane (or F5-brane, for short) \cite{GSflux} in the
context of Kaluza--Klein reductions.  Starting with eleven-dimensional
Minkowski space, the F5-brane is obtained by considering a
Kaluza--Klein reduction along the orbits generated by a Killing vector
consisting of a translation and a rotation.  The reduction will
preserve some supersymmetry if the rotation belongs to the isotropy
algebra of some nonzero Killing spinor of the Minkowski vacuum.  This
suggests considering the Kaluza--Klein reduction by the most general
Killing vector in Minkowski space subject to some natural conditions,
namely that the quotient be a smooth lorentzian manifold.  This
problem has been addressed in \cite{DGGH3}, although there no
particular attention has been paid to supersymmetric fluxbranes or to
Lorentz transformations other than spatial rotations.  The space of
Killing vectors is of course isomorphic to the Poincaré algebra, and
the reductions satisfying these conditions form a subset of this
algebra.  Generic points in this parameter space will break
supersymmetry completely, but there are special loci for which
supersymmetry is preserved.  In this paper we determine these
loci.  There are two families of solutions intersecting in a common
two-parameter family.

Any translation preserves supersymmetry, but a Lorentz transformation
which does must belong to the isotropy algebra of some spinor.  In
dimension greater than two, there are three types of (pure) Lorentz
transformations: boosts, rotations and null rotations.  Only null
rotations and some ``special'' rotations preserve spinors.  Reductions
involving these special rotations generate type IIA configurations
which can be interpreted as flux $p$-branes (or $\text{F}p$-branes,
for short) for $p=1,3,5$, but we also have the possibility of reducing
along null rotations. This gives rise to configurations which we call
\emph{nullbranes} (or N$p$-branes) as they have a null RR 2-form field
strength.  Whereas supersymmetric fluxbranes can be interpreted as
intersections of F7-branes, nullbranes can be thought of infinitely
boosted F7-branes\footnote{JS would like to thank M.~Berkooz for
  suggesting this interpretation.}, even though their physical
interpretation is not clear to us at the present time, these being
time dependent backgrounds in string theory preserving one half of the
spacetime supersymmetry.  Reductions along linear combinations of the 
Killing vectors giving rise to the flux/null-branes are also possible. 
The type IIA configurations obtained in this way interpolate continuously 
between the fluxbranes and nullbranes discussed before.

The formulation developed in this paper allows an immediate extension
to many other supersymmetric M-theory backgrounds.  This is discussed
and applied to M-brane backgrounds and their intersections in a
separate article \cite{FigSimBranes}, in which we construct
supersymmetric composite configurations of type IIA/B branes
(D-branes, fundamental strings, NS5-branes, waves, etc) and
flux/null-branes.  In a second forthcoming article we consider
supersymmetric reductions of the maximally supersymmetric M-theory
backgrounds of AdS type \cite{FigSimAdS}.

There remain many interesting open questions regarding F-branes such
as which D-brane configurations are possible in these backgrounds (or
their U-duals) and which is the corresponding field theory one can
define on them in certain decoupling limits.  Furthermore, while
standard D-branes admit an open string description on flat space, the
existence of such a description for F-branes remains an open question.

This paper is organised as follows.  In Section~\ref{sec:KK} we
discuss the geometric context of the present paper: namely the
supersymmetric Kaluza--Klein reduction of the flat M-theory vacuum and
we classify those reductions resulting in supersymmetric smooth IIA
backgrounds, what we call (smooth, generalised) supersymmetric IIA
fluxbranes.  We will see that there are two families of generalised
supersymmetric fluxbranes: one family contains the supersymmetric
fluxbranes recently discussed in \cite{GSflux,Uranga,RTFlux} and in
addition a novel $\tfrac14$-BPS fluxstring solution, and the other
family contains new solutions called here nullbranes as well as new
solutions which interpolate between them and some of the known
fluxbranes.  We can summarise the possible generalised supersymmetric
fluxbranes as follows.  As elementary objects, we have a $\half$-BPS
F5-brane, a $\half$-BPS N7-brane, a $\tfrac14$-BPS F3-brane, a
$\tfrac14$-BPS F1-brane and an $\tfrac18$-BPS F1-brane.  We also find
interpolating solutions (which we call generalised fluxbranes): a
$\tfrac14$-BPS solution interpolating between the F5- and the
N7-branes and a $\tfrac18$-BPS solution interpolating between the F3-
and the N7-branes.  All these geometries are constructed explicitly in 
Section~\ref{sec:fluxbranes}.
Finally Section~\ref{sec:dualities} contains a comprehensive analysis
of the U-dual configurations which can be obtained from the solutions
constructed in Section~\ref{sec:fluxbranes}.  This is illustrated with
the aid of diagrams showing the U-duality orbits of some of the
elementary fluxbranes.

\textsc{Note added.}  There is some overlap between the results in
this paper and those in \cite{RTFlux}, which appeared as we were
completing the present work.

\section{Kaluza--Klein reduction and generalised fluxbranes}
\label{sec:KK}

We start by considering Kaluza--Klein reductions of the maximally
supersymmetric vacuum of M-theory described by eleven-dimensional
Minkowski space, and in this way classify all the smooth
supersymmetric solutions of type IIA supergravity which are
Kaluza--Klein reductions of the M-theory vacuum.  These geometries
will be described explicitly in the next section.

\subsection{The geometrical setting}

The general geometrical setting is the following (see also
\cite{DGGH3}).  We will consider a one-parameter subgroup of the group
of isometries of Minkowski space, in other words a one-dimensional
subgroup $G$ of the Poincaré group acting on Minkowski space.
Topologically, $G$ is either a circle or a line.  In traditional
treatments of Kaluza--Klein reduction, $G$ is always taken to be a
circle subgroup; but circles do not act freely on Minkowski space.  It
is possible to follow tradition and demand that $G$ be a circle; but
in order to have a smooth quotient, one is forced to introduce
``identifications'' in Minkowski space which effectively close the
integral curves of the Killing vector generating the action.  This
practice is standard in the context of fluxbranes, and although it has
its merits, we prefer not to follow it here.  To restate, all our
solutions will be smooth quotients of Minkowski space by the action of
$G\cong \RR$.  If desired, these solutions \emph{could} be viewed as
Kaluza--Klein reductions on a circle.  Indeed, quotienting by $\RR$
can be done in two steps: quotienting by $\ZZ$ (i.e., making
identifications in Minkowski space) and then quotienting by the circle
$\RR/\ZZ$.  The action of $\ZZ$ introduces a length scale in the
problem, which is ($2\pi$ times) the radius of the circle.

If the $G$-action is free and has spacelike orbits\footnote{Since we
  want to induce a metric in the space of orbits, it is important that
  the Killing vector never be null; hence it has to be always
  spacelike or always timelike.  Since we are interested in
  constructing solutions of type II supergravity in signature (9,1) we
  restrict ourselves to spacelike orbits.  Nothing prevents us from
  considering timelike orbits and in this way obtain fluxbrane
  solutions of euclidean supergravity theories of the type considered
  in \cite{HullTimeLike}, but this will be left for another time.},
the ten-dimensional space obtained by Kaluza--Klein reduction will be
a solution of the type IIA supergravity.  In addition, this solution
will be supersymmetric provided that $G$ leaves some spinor invariant.
Let $\xi$ be the Killing vector generating the $G$ action
infinitesimally.  The action will be (locally) free if and only if
$\xi$ is nowhere-vanishing.  With our choice of (mostly plus) metric,
demanding that the norm $\|\xi\|^2$ be everywhere positive guarantees
that the orbits are spacelike.  Under further mild restrictions
(namely that every point in Minkowski space should have trivial
stabiliser) the space of orbits will be a smooth manifold inheriting a
lorentzian metric.

In adapted coordinates, where $\xi= \d_z$, we can write the
eleven-dimensional Minkowski metric as
\begin{equation}
  \label{eq:kkstring}
  ds^2(\EE^{10,1}) = e^{-2\phi/3} g + e^{4\phi/3} (dz + A)^2~,
\end{equation}
where $g$ is the ten-dimensional metric in the string frame, $\phi$ is
the dilaton and $A$ is the RR $1$-form potential of type IIA
supergravity.  By construction, the triple $(g,\phi,A)$ will satisfy
the equations of motion of IIA supergravity, provided that we set the
other field strengths $F_3$ and $F_4$ in the theory to zero.  This
solution will be supersymmetric if and only if $\xi$ preserves some
Killing spinors.  In the case where we make identifications in
Minkowski space in order to view this as a circle reduction, it is
convenient to introduce a length scale $R$, the radius of the circle,
and then write $z = R\chi$, where $\chi$ is an angular variable taking
values in $\RR/2\pi\ZZ$.  In this case, it is also convenient to
rescale the RR 1-form potential, so that the metric becomes the more
familiar
\begin{equation}
  ds^2(\EE^{10,1}) = e^{-2\phi/3} g + e^{4\phi/3} R^2 (d\chi +
  C_{(1)})^2~.
\end{equation}

The Killing vector $\xi$ acts on a Killing spinor $\varepsilon$ via
the spinorial Lie derivative (see, e.g., \cite{Kosmann} and also
\cite{JMFKilling}), defined by
\begin{equation}
  L_\xi \varepsilon = \nabla_\xi \varepsilon + \tfrac14 \nabla_a \xi_b
  \Gamma^{ab} \varepsilon~.
\end{equation}
Since Killing spinors are parallel in the Minkowski vacuum, the
condition that $\xi$ preserves supersymmetry becomes the algebraic
condition that, for some nonzero parallel spinor $\varepsilon$,
\begin{equation}
  \label{eq:supersymmetry}
  \tfrac14 \nabla_a \xi_b \Gamma^{ab} \varepsilon = 0~.
\end{equation}
Alternatively, relative to adapted coordinates, a spinor is invariant
under $\xi=\d_z$ if and only if it does not depend explicitly on
$z$.

The most general Killing vector in Minkowski space is the sum of a
translation and a Lorentz transformation:
\begin{equation}
  \xi = a^\mu \d_\mu + \half \omega^{\mu\nu} (x_\mu \d_\nu - x_\nu
  \d_\mu)~,
\end{equation}
where $\omega^{\mu\nu} = - \omega^{\nu\mu}$, and the condition
\eqref{eq:supersymmetry} guaranteeing the preservation of
supersymmetry becomes
\begin{equation}
 \label{eq:susypres}
  \tfrac14 \omega^{\mu\nu} \Sigma_{\mu\nu} \varepsilon = 0~.
\end{equation}
In other words, we see that the translation can be arbitrary and that
the Lorentz transformation must belong to the isotropy algebra of a
spinor.

For the purposes of this paper, by a \emph{(generalised) fluxbrane} we
will mean a smooth solution of IIA supergravity obtained as the
Kaluza--Klein reduction of eleven-dimensional Minkowski space.  In
what follows we will classify these and construct two multiparameter
families of generalised supersymmetric fluxbranes containing as special
cases all the supersymmetric fluxbranes which have been hitherto
considered in the literature.

\subsection{Classification of smooth supersymmetric fluxbranes}

Let us outline the mathematical problem of classifying supersymmetric
generalised fluxbranes (of type IIA supergravity).  This problem
reduces to finding free actions of a group $G \cong \RR$ on Minkowski
space preserving the metric, leaving some parallel spinors invariant
and possessing spacelike orbits.  In other words, we are interested in
classifying one-parameter subgroups of the spinor isotropy groups of
(the spin cover of) the Poincaré group, which act freely on Minkowski
space with spacelike orbits, up to conjugation.

It is more convenient to work in terms of the Lie algebra.   Killing
vectors in Minkowski space are in one-to-one correspondence with the
Poincaré algebra $\fp = \fl \ltimes \ft \cong \fso(10,1) \ltimes
\RR^{10,1}$, where $\fl$ is the Lorentz subalgebra and $\ft$ is the
translation ideal.  Let $\fm \subset \fp$ denote the \emph{subset}
corresponding to Killing vectors $\xi$ which obey the following
conditions
\begin{enumerate}
\item $\|\xi\|^2 = g(\xi,\xi) > 0$; and
\item $\xi$ preserves some spinors.
\end{enumerate}
Notice that if $\xi\in\fm$ then $s\xi\in\fm$ for any nonzero real
number $s$.  Although there is physics in the scale $s$, it is
convenient for classification purposes to identify any two collinear
Killing vectors, as they have the same orbits, but parametrised
differently.  We therefore introduce the \emph{moduli space} $\eM$
of supersymmetric Kaluza--Klein reductions of flat space as the
real projective space of the subset $\fm$.  We now proceed to
determine $\eM$.

Let $\xi$ be a Killing vector and let us write it uniquely as
\begin{equation*}
  \xi = \tau + \lambda~,
\end{equation*}
where $\tau\in\ft$ is a translation and $\lambda\in\fl$ is a Lorentz
transformation.  The translation component cannot be zero, since every
Lorentz transformation fixes a point (the ``origin'') and hence
vanishes there.  Moreover, since $\|\xi\|^2 = \|\tau\|^2$ at the
origin, condition (1) above says that $\tau$ must be spacelike at the
origin, and hence everywhere.  Our strategy to determine the moduli
space $\eM$ will be the following: we will find a normal form for
$\xi$ exploiting the freedom to conjugate by the Poincaré group, and
then impose the conditions on the normal form.

By conjugating with a Lorentz transformation we can bring $\lambda$ to
a normal form.  Which form depends on the type of element it is.  A
generic $\lambda \in \fso(10,1)$ fixes one direction.  This direction
can be spacelike, timelike or null.  Accordingly, $\lambda$ can take
one of the following three normal forms:
\begin{itemize}
\item[] (spacelike)
  \begin{equation}
    \label{eq:sNF}
    \lambda = B_{01}(\gamma) + R_{23}(\beta_1) + R_{45}(\beta_2) +
    R_{67}(\beta_3) + R_{89}(\beta_4)~;
  \end{equation}
\item[] (timelike)
  \begin{equation}
    \label{eq:tNF}
    \lambda = R_{12}(\beta_1) + R_{34}(\beta_2) + R_{56}(\beta_3) +
    R_{78}(\beta_4) + R_{9\natural}(\beta_5)~;
  \end{equation}
\item[] (null)
  \begin{equation}
    \label{eq:nNF}
    \lambda = N_{+1}(u) + R_{23}(\beta_1) + R_{45}(\beta_2) +
    R_{67}(\beta_3) + R_{89}(\beta_4)~,
  \end{equation}
\end{itemize}
where $\pm = \natural \pm 0$ with $\natural$ the tenth spacelike
direction, $R_{ij}(\beta)$ is an infinitesimal rotation with parameter
$\beta$ in the $(ij)$ plane, $B_{0i}(\gamma)$ is an infinitesimal
boost with parameter $\gamma$ along the $i$th direction and
$N_{+i}(u)$ is a null rotation with parameter $u$ in the $i$th
direction.

Bringing $\lambda$ to one of these normal forms does not use up all
the freedom of conjugation, since we can still conjugate by those
elements of the Poincaré group which fix the normal form.  All normal
forms are fixed under conjugation by translations, which corresponds
to changing the origin in Minkowski space. It is possible to change
the origin in order the bring the translation $\tau$ to normal form.
This normal form depends on $\lambda$.  Under a change of origin,
\begin{equation*}
  \xi = \tau + \lambda \mapsto \tau + [\lambda,\tau'] + \lambda~,
\end{equation*}
whence we can, by choosing the origin appropriately, get rid of any
component of $\tau$ in the image of $[\lambda,-]$.

If $\lambda$ has normal form \eqref{eq:sNF}, by a change of origin and
a rescaling (which we are allowed to do since the moduli space $\eM$
is projective) we can always choose $\tau = \d_\natural$.  In
particular this means that as vector fields in Minkowski space, $\tau$
and $\lambda$ are orthogonal, whence
\begin{equation*}
  \|\xi\|^2 = \|\tau\|^2 + \|\lambda\|^2~.
\end{equation*}
It is easy to see that if the boost parameter $\gamma\neq 0$ then
there are points in Minkowski space where $\xi$ is not spacelike:
simply take $|x^1|$ large enough.  Therefore for spacelike orbits, we
require $\gamma = 0$, whence $\lambda$ also fixes a timelike
direction, so it is a special case of the normal form \eqref{eq:tNF}.

If $\lambda$ has normal form \eqref{eq:tNF} and all $\beta$'s are
different from zero, there exists a choice of origin in which $\tau$
is timelike, violating one of the above conditions.  Therefore at most
four $\beta$'s can be nonzero and hence there exist coordinates in
which, after rescaling, $\xi = \d_\natural + \lambda$, where $\lambda$
takes the form
\begin{equation*}
  \lambda = R_{12}(\beta_1) + R_{34}(\beta_2) + R_{56}(\beta_3) +
    R_{78}(\beta_4) \in \fso(8)~.
\end{equation*}
Now, supersymmetry never constrains the translation component, but it
does impose constraints on $\lambda$.  Here $\lambda \in \fso(8)$
preserves a spinor if and only if it belongs to a $\fspin(7)$
subalgebra.  By permuting the coordinates if necessary, this condition
translates into the vanishing of the sum of the $\beta$'s.  This gives
rise to a three-parameter family of vector fields
\begin{equation}
  \label{eq:fluxbranes}
  \xi = \d_\natural + R_{12}(\beta_1) + R_{34}(\beta_2) +
  R_{56}(\beta_3) + R_{78}(\beta_4)~,
\end{equation}
with $\beta_1 + \beta_2 + \beta_3 + \beta_4 = 0$.

Finally if $\lambda$ has normal form \eqref{eq:nNF}, we must again
have some parameter among $u,\beta_i$ vanishing, for otherwise by a
suitable change of origin we could set $\tau = a \d_-$, violating the
condition that $\tau$ be spacelike.  Therefore either $u$ or one of
the $\beta$'s must vanish.  If $u=0$ we are back in the case treated
previously.  Therefore let us assume that $u\neq 0$ and that
$\beta_4 =0$, say.  By choosing the origin suitably we can write $\xi
= \tau + \lambda$, where
\begin{footnotesize}
\begin{equation*}
  \tau = a \d_- + b \d_8 + c \d_9 \quad\text{and}\quad
  \lambda =  N_{+1}(u) + R_{23}(\beta_1) + R_{45}(\beta_2) +
    R_{67}(\beta_3)~.
\end{equation*}
\end{footnotesize}
The condition that $\tau$ be spacelike forces either $b$ or $c$ to be
nonzero.  Moreover the condition that $\xi$ be spacelike forces
$a=0$.  To see this simply notice that
\begin{equation*}
  \begin{split}
    \|\xi\|^2 &= \|\tau\|^2 + \|\lambda\|^2 + 2 g(\tau,\lambda)\\
    &= \|\tau\|^2 + \|\lambda\|^2 - 2 a u x^1~,
  \end{split}
\end{equation*}
where $\|\tau\|^2>0$, $\|\lambda\|^2 \geq 0$ are independent of $x^1$
and hence if we take $|x^1|$ large enough we eventually find that
$\|\xi\|^2$ is not positive.  Therefore in some coordinates, and after
rescaling,
\begin{equation*}
  \xi = \d_9 + N_{+1}(u) + R_{23}(\beta_1) + R_{45}(\beta_2) +
    R_{67}(\beta_3)~.
\end{equation*}
It remains to impose that $\xi$ preserves some spinors.  This means
that for some nonzero spinor $\varepsilon$, its Lie derivative $L_\xi
\varepsilon$ along $\xi$ vanishes.  Using \eqref{eq:susypres} this is
equivalent to the algebraic equation
\begin{equation*}
  \left( u \Sigma_{+1} + \beta_1 \Sigma_{23} + \beta_2 \Sigma_{45} +
  \beta_3 \Sigma_{67} \right) \varepsilon = 0~.
\end{equation*}
To solve it, notice that this equation is of the form
\begin{equation}
  \label{eq:decomp}
  (N + S) \varepsilon = 0~,
\end{equation}
where $N = u \Sigma_{+1}$ is nilpotent (in fact, $N^2=0$) and $S =
\beta_1 \Sigma_{23} + \beta_2 \Sigma_{45} + \beta_3 \Sigma_{67}$ is
semisimple and moreover $N$ and $S$ commute.  As a result,
\eqref{eq:decomp} is true if and only if
\begin{equation*}
  N \varepsilon = 0 \qquad \text{and}\qquad S \varepsilon = 0~.
\end{equation*}
The first equation says that $\Gamma_+\varepsilon = 0$ and the second
equation says that $S \in \fso(6)$ is actually in an $\fsu(3)$
subalgebra.  Again, up to a permutation of the coordinates, this means
that the sum of the $\beta$'s vanishes.

We can summarise the above results as follows. There are two families
of spacelike Killing vectors in Minkowski space which induce
supersymmetric Kaluza--Klein reductions.  There exists a coordinate
system $(z, y^i, y^\pm)$, with $i=1,\dots,8$, where the metric takes
the form
\begin{equation}
  \label{eq:metric}
  ds^2(\EE^{10,1}) = 2 dy^+ dy^- + \sum_{i=1}^8 dy^i dy^i + dz^2
\end{equation}
and where, to an overall scale, the two families of Killing vectors
are given by
\begin{equation}
  \label{eq:fluxfamily}
  \xi = \d_z + R_{12}(\beta_1) + R_{34}(\beta_2) + R_{56}(\beta_3) +
  R_{78}(\beta_4)~,\qquad \sum_i \beta_i = 0~,
\end{equation}
and by
\begin{equation}
  \label{eq:nullfamily}
  \xi = \d_z + N_{+1}(u) + R_{34}(\beta'_1) + R_{56}(\beta'_2) +
  R_{78}(\beta'_3)~,\qquad \sum_i \beta'_i = 0~.
\end{equation}
We should point out that the parameter $u$ is ineffective; that is, it
is only important to distinguish between two cases $u=0$ and $u\neq
0$.  In this latter case, we can set $u$ to any desired nonzero value
by rescaling $y^\pm \mapsto c^{\pm 1} y^\pm$ for a suitable $c$.  Such
rescalings are of course Lorentz transformations.

To complete the proof that these vector fields give rise to smooth
reductions, we must ensure that the integrated action is free: meaning
that no point is left fixed and every point has trivial stabiliser.
Indeed, the action on Minkowski space of the element $\exp t \xi$ in
the subgroup $G\cong\RR$ generated by $\xi$ takes the form
\begin{equation}
 (z, y^\mu) \mapsto (z + t, \Lambda(t)^\mu{}_\nu y^\nu)~,
\end{equation}
whence the action is manifestly free.  This means that the space of
orbits $\EE^{10,1}/G$ is smooth.  Let us now describe the explicit
geometry of these supersymmetric fluxbranes.

\section{Generalised supersymmetric fluxbranes}
\label{sec:fluxbranes}

To best describe the explicit geometry of the generalised
supersymmetric fluxbranes obtained by reducing Minkowski space along
the orbits of the Killing vectors given by equations
\eqref{eq:fluxfamily} and \eqref{eq:nullfamily}, we will work in
coordinates adapted to the Killing vector.  This turns out to be very
easy, once we observe that $\xi$ is simply a dressed version of its
translation component:
\begin{equation}
  \xi = U \d_z U^{-1} \qquad\text{where}\quad U = \exp\left( -z
  \lambda \right)~.
\end{equation}
Let us introduce coordinates $(x^i, x^\pm)$ related to the
coordinates $(y^i, y^\pm)$ by
\begin{equation}
  x = U\, y~.
\end{equation}
It follows easily that $\xi x = 0$, so that $x$ are good coordinates
for the space of orbits.  Since $\lambda$ is an infinitesimal Lorentz
transformation, the $x$ are linearly related to the $y$ with
$z$-dependent coefficients.  Indeed, let us denote by $B$ the constant
$10 \times 10$ real matrix such that
\begin{equation}
  \lambda y = B y~.
\end{equation}
Therefore the new coordinates are given by
\begin{equation}
  x = \exp\left( -z B \right) y~,
\end{equation}
where $\exp$ here means the matrix exponential.  For $\lambda$ in
either of the normal forms described at the end of the last section,
it is very easy to write an explicit expression for this matrix
exponential; but we will refrain from doing so here, as it is not
needed in order to write down the IIA background.

Indeed, it is a simple matter to rewrite the metric \eqref{eq:metric}
in terms of the new variables, obtaining
\begin{equation}
  ds^2(\EE^{10,1}) = \Lambda (dz + A)^2 + dx^\dagger
  \left( 1 - \Lambda^{-1} (Bx)(Bx)^\dagger \right) dx~,
\end{equation}
where $v^\dagger = v^t \eta$ is the adjoint relative to the Minkowski
metric, and
\begin{equation}
  \label{eq:fluxLA}
  \Lambda = 1 + (Bx)^\dagger (Bx) \qquad \text{and} \qquad
  A = \Lambda^{-1} (Bx)^\dagger dx~.
\end{equation}
Using the Kaluza--Klein ansatz \eqref{eq:kkstring} we can read off the
IIA background which has $F_3=F_4=0$ and in the string frame the
nontrivial fields are given by
\begin{equation}
  \label{eq:fluxbrane}
  \phi = \tfrac34 \log \Lambda \qquad\text{and}\qquad
  g = \Lambda^{1/2} dx^\dagger \left( 1 - \Lambda^{-1}
  (Bx)(Bx)^\dagger \right) dx~,
\end{equation}
together with $A$ in the previous equation, which also contains the
definition of $\Lambda$.  The parameters in this solution are hidden
in $B$, which is the matrix of an infinitesimal Lorentz transformation
of the form described in equations \eqref{eq:fluxfamily} or
\eqref{eq:nullfamily}.  As discussed in the previous section, we can
always choose the $x$ coordinates in such a way that $B$ only depends
on $3$ real parameters which are unconstrained; although in one of the
cases the parameter is ineffective.  We keep it because we can in this
way discuss the limit $u\to 0$.  Some subvarieties of the moduli space
of supersymmetric fluxbranes are already known, as we now discuss.

To help comparison with the literature, let us write down the matrix
$B$ explicitly relative to the basis $\{x^i, x^+, x^-\}$:
\begin{equation}
  \label{eq:Bmatrix}
  B= 
  \begin{pmatrix}
    0 & -\beta_1 & 0 & 0 & 0 & 0 & 0 & 0 & 0 & u \\
    \beta_1  & 0 & 0 & 0 & 0 & 0 & 0 & 0 & 0 & 0 \\
    0 & 0 & 0 & -\beta_2 & 0 & 0 & 0 & 0 & 0 & 0 \\
    0 & 0 & \beta_2  & 0 & 0 & 0 & 0 & 0 & 0 & 0 \\
    0 & 0 & 0 & 0 & 0 & -\beta_3 & 0 & 0 & 0 & 0 \\
    0 & 0 & 0 & 0 & \beta_3  & 0 & 0 & 0 & 0 & 0 \\
    0 & 0 & 0 & 0 & 0 & 0 & 0 & -\beta_4 & 0 & 0 \\
    0 & 0 & 0 & 0 & 0 & 0 & \beta_4  & 0 & 0 & 0 \\
    - u & 0 & 0 & 0 & 0 & 0 & 0 & 0 & 0 & 0 \\
    0 & 0 & 0 & 0 & 0 & 0 & 0 & 0 & 0 & 0
  \end{pmatrix}
\end{equation}
where either $u=0$ or $\beta_1=0$ (or both) and where $\beta_1 + 
\beta_2 + \beta_3 + \beta_4 = 0$.  We will now consider several
special cases.

\subsection{Supersymmetric fluxbranes}

If $u = 0$, so that the Lorentz transformation is purely a rotation,
we obtain, in addition to the standard supersymmetric fluxbranes
considered previously in \cite{GSflux,Uranga,RTFlux}, a novel
fluxstring configuration preserving $1/4$ of the supersymmetry.  The
solutions described here correspond to the supersymmetric points in
the space of solutions described in \cite{DGGH3}; although that work
did not address the question of supersymmetry.

We have several possibilities depending on how many of the $\beta$'s
are zero.  If three $\beta$'s vanish, then supersymmetry implies that
all $\beta$'s must vanish and the resulting configuration is the type
IIA vacuum.  Therefore, if we are to obtain a nontrivial
supersymmetric fluxbrane, at most two $\beta$'s can be zero.

\subsubsection{The supersymmetric F5-brane revisited}

If $\beta_1=\beta_2=0$, and hence $\beta_3=-\beta_4 = \beta$, we
recover the supersymmetric F5-brane of \cite{GSflux}.  The
matrix $B$ corresponds to an element in the Cartan subalgebra of
$\fsp(1) \subset \fso(4) \subset \fso(9,1)$:
\begin{equation}
  \label{eq:F5matrix}
  B= 
  \begin{pmatrix}
    0 & 0 & 0 & 0 & 0 & 0 & 0 & 0 & 0 & 0 \\
    0 & 0 & 0 & 0 & 0 & 0 & 0 & 0 & 0 & 0 \\
    0 & 0 & 0 & 0 & 0 & 0 & 0 & 0 & 0 & 0 \\
    0 & 0 & 0 & 0 & 0 & 0 & 0 & 0 & 0 & 0 \\
    0 & 0 & 0 & 0 & 0 & -\beta & 0 & 0 & 0 & 0 \\
    0 & 0 & 0 & 0 & \beta & 0 & 0 & 0 & 0 & 0 \\
    0 & 0 & 0 & 0 & 0 & 0 & 0 & \beta & 0 & 0 \\
    0 & 0 & 0 & 0 & 0 & 0 & -\beta & 0 & 0 & 0 \\
    0 & 0 & 0 & 0 & 0 & 0 & 0 & 0 & 0 & 0 \\
    0 & 0 & 0 & 0 & 0 & 0 & 0 & 0 & 0 & 0
  \end{pmatrix}
\end{equation}
Here $\beta$ is the parameter related to the magnetic field in
\cite{GSflux}.  The solution is given explicitly by
\eqref{eq:fluxbrane} with
\begin{equation}
  \Lambda = 1 + \beta^2 r^2 \quad \text{and} \quad
  A = \frac{\beta}{1+\beta^2 r^2} \left( x^5 dx^6 - x^6 dx^5 - x^7
    dx^8 + x^8 dx^7\right)~,
\end{equation}
where $r$ is the radial coordinate in the $4$-plane spanned by the
coordinates $x^5,\dots,x^8$ transverse to the fluxbrane.  We can
rewrite this solution in a more familiar form, by introducing
coordinates:
\begin{equation}
  \label{eq:F5coords}
  x^5 + i x^6 = r \cos\theta e^{i(\psi + \varphi)} \qquad \text{and}
  \qquad x^7 + i x^8 = r \sin\theta e^{i(\psi - \varphi)}~,
\end{equation}
where $\theta \in [0,\frac\pi2]$, $\psi\in [0,\pi]$ and
$\varphi\in \RR/2\pi\ZZ$.  In terms of the new coordinates,
\begin{equation}
  \label{eq:F5RR1form}
  A = \frac{\beta r^2}{1+\beta^2 r^2} \left( d\varphi + \cos(2\theta)
  d\psi\right)~,
\end{equation}
and the metric becomes
\begin{multline}
  \label{eq:F5metric}
  g = \Lambda^{1/2} \left[ ds^2(\EE^{5,1}) + dr^2 + r^2 d\theta^2
    + r^2 \sin^2(2\theta)d\psi^2 \right]\\
  + \Lambda^{-1/2} r^2 \left(d\varphi + \cos 2\theta d\psi\right)^2~.
\end{multline}

This solution represents a F5-brane.  Indeed, the IIA RR $2$-form
field $F=dA$ in this solution has a nontrivial ``charge'', as can be
seen from the (normalised) integral of $F \wedge F$ on the transverse
$\EE^4$:
\begin{equation}
  \frac{1}{8\pi^2} \int_{\EE^4} F\wedge F = \lim_{\rho \to \infty}
  \frac{1}{8\pi^2} \int_{r\leq \rho} F \wedge F = \lim_{\rho\to
    \infty} \frac{1}{8\pi^2} \int_{r = \rho} A \wedge F =
    \frac{1}{\beta^2}~,
\end{equation}
where we have used that the orientation on the 3-sphere $r=\rho$
induced by the natural orientation $dx^5 \wedge dx^6 \wedge dx^7
\wedge dx^8$ on $\RR^4$ agrees with $d\theta \wedge d\varphi \wedge
d\psi$.  The F5-brane preserves one half of the supersymmetry of
the eleven-dimensional vacuum. 

The near-horizon geometry ($r\to 0$) of the F5-brane is
flat. Indeed the metric is asymptotic to $\EE^{5,1}$ times a cone
metric:
\begin{equation}
 dr^2 + r^2 \left[ d\theta^2 + \sin^2(2\theta) d\psi^2 +
 (d\varphi+\cos(2\theta)d\psi)^2 \right]~.
\end{equation}
The metric of the base of the cone (that is, the quantity in square
brackets) is that of the round $3$-sphere, whence the cone metric is
that of $\EE^4$.  In the other limit ($r\to\infty$) we obtain a
conformally cylindrical geometry.  Indeed, $h$ is asymptotically
conformal to a metric which is a product of $\EE^{5,1}$ with
\begin{equation}
 dr^2 + r^2 \left[ d\theta^2 + \sin^2(2\theta) d\psi^2  \right] +
 \frac{1}{\beta^2} (d\varphi+\cos(2\theta)d\psi)^2~.
\end{equation}
This is the metric on the total space of a circle bundle over $\RR^3$
where the radius of the circle is constant and equal to $1/\beta$.

\subsubsection{The supersymmetric F3-brane}

If $\beta_1=0$, but the rest of the $\beta$'s are different from zero
(but add up to zero) we obtain the supersymmetric flux-threebrane
discussed in \cite{Uranga,RTFlux}.  In this case the matrix $B$ is a
Cartan subalgebra of $\fsu(3) \subset \fso(6) \subset \fso(9,1)$:
\begin{equation}
  \label{eq:F3matrix}
  B= 
  \begin{pmatrix}
    0 & 0 & 0 & 0 & 0 & 0 & 0 & 0 & 0 & 0 \\
    0 & 0 & 0 & 0 & 0 & 0 & 0 & 0 & 0 & 0 \\
    0 & 0 & 0 & -\beta_2 & 0 & 0 & 0 & 0 & 0 & 0 \\
    0 & 0 & \beta_2 & 0 & 0 & 0 & 0 & 0 & 0 & 0 \\
    0 & 0 & 0 & 0 & 0 & -\beta_3 & 0 & 0 & 0 & 0 \\
    0 & 0 & 0 & 0 & \beta_3 & 0 & 0 & 0 & 0 & 0 \\
    0 & 0 & 0 & 0 & 0 & 0 & 0 & -\beta_4 & 0 & 0 \\
    0 & 0 & 0 & 0 & 0 & 0 & \beta_4 & 0 & 0 & 0 \\
    0 & 0 & 0 & 0 & 0 & 0 & 0 & 0 & 0 & 0 \\
    0 & 0 & 0 & 0 & 0 & 0 & 0 & 0 & 0 & 0
  \end{pmatrix}
\end{equation}
The solution is again given explicitly by \eqref{eq:fluxbrane} with
\begin{equation}
  \Lambda = 1 + \sum_{i=2}^4\beta_i^2 \vert z_i \vert^2 \quad
  \text{and} \quad A = \frac{\Lambda^{-1}}{2i}
  \sum_{i=2}^4\beta_i\left(\bar{z}_idz_i - z_i d\bar{z}_i\right)~,
\end{equation}
where $z_i$, $i=2,3,4$, correspond to complex coordinates in the
6-plane spanned by the real coordinates $x^3,\dots ,x^8$ transverse to
the fluxbrane.  We can rewrite this in terms of the radial distance to
the fluxbrane by introducing coordinates:
\begin{equation}
 \begin{aligned}[m]
  \label{eq:F3coords} 
   x^3+ix^4&=z_1=r\cos\theta_1 e^{i\varphi_1}\\
   x^5+ix^6&=z_2=r\sin\theta_1\cos\theta_2 e^{i\varphi_2}\\
   x^7+ix^8&=z_3=r\sin\theta_1 \sin\theta_2e^{i\varphi_3}~,
 \end{aligned}
\end{equation}
where $\theta_{1,2}\in [0,\frac{\pi}{2}]$ and $\varphi_i \in
\RR/2\pi\ZZ$ for all $i$.  In terms of the new coordinates, the scalar
function becomes
\begin{equation}
 \label{eq:F3lambda}
 \Lambda= 1 + r^2\left(\beta_2^2\cos^2\theta_1 +
 \beta_3^2\sin^2\theta_1 \cos^2\theta_2 +
 \beta_4^2\sin^2\theta_1\sin^2 \theta_2\right)~,
\end{equation}
whereas the RR 1-form potential is given by
\begin{footnotesize}
\begin{equation}
  \label{eq:F3RR1form}
  A = \Lambda^{-1}r^2\left( \beta_2 \cos^2\theta_1 d\varphi_1 +
  \beta_3 \sin^2\theta_1\cos^2\theta_2 d\varphi_2 + 
  \beta_4\sin^2\theta_1\sin^2 \theta_2 d\varphi_3\right)~,
\end{equation}
\end{footnotesize}
and the metric becomes
\begin{footnotesize}
\begin{multline}
  \label{eq:F3metric}
  g = \Lambda^{1/2} \left[ ds^2(\EE^{3,1}) + dr^2 + r^2
    \left(d\theta_1^2 + \sin^2 \theta_1 d\theta_2^2\right) \right. \\
  \left. + r^2\left(\cos^2\theta_1 d\varphi_1^2 +
      \sin^2\theta_1\cos^2\theta_2 d\varphi_2^2 +
      \sin^2\theta_1\sin^2\theta_2 d\varphi_3^2\right) \right]\\
  - \Lambda^{-1/2} r^4 \left(\beta_2 \cos^2\theta_1 d\varphi_1 +
    \beta_3 \sin^2\theta_1\cos^2\theta_2 d\varphi_2 +
    \beta_4\sin^2\theta_1\sin^2 \theta_2 d\varphi_3\right)^2~.
\end{multline}
\end{footnotesize}
The solution preserves $1/4$ of the supersymmetry.

\subsubsection{Supersymmetric fluxstrings}

If none of the $\beta$'s vanish, there are two possibilities.  Either
the $\beta$'s add up to zero pairwise or they do not.  In the former
case, the matrix $B$ belongs to a Cartan subalgebra of $\fsp(1) \oplus
\fsp(1) \subset \fso(4) \oplus \fso(4) \subset \fso(8) \subset
\fso(9,1)$, whereas in the latter, it belongs to a Cartan subalgebra
of $\fsu(4) \subset \fso(8) \subset \fso(9,1)$.  In either case we
have a supersymmetric fluxstring.  In the former case the fluxstring
preserves $1/4$ of the supersymmetry, whereas in the latter case it
preserves $1/8$.  The latter case was discussed in
\cite{Uranga,RTFlux}, the former case seems new.

Let us examine, first of all, the case when $B$ belongs to a Cartan 
subalgebra of $\fsp(1) \oplus \fsp(1) \subset \fso(4) \oplus \fso(4) 
\subset \fso(8) \subset \fso(9,1)$:
\begin{equation}
%  \label{eq:F1amatrix}
  B= 
  \begin{pmatrix}
    0 & -\beta & 0 & 0 & 0 & 0 & 0 & 0 & 0 & 0 \\
    \beta & 0 & 0 & 0 & 0 & 0 & 0 & 0 & 0 & 0 \\
    0 & 0 & 0 & \beta & 0 & 0 & 0 & 0 & 0 & 0 \\
    0 & 0 & -\beta & 0 & 0 & 0 & 0 & 0 & 0 & 0 \\
    0 & 0 & 0 & 0 & 0 & -\tilde{\beta} & 0 & 0 & 0 & 0 \\
    0 & 0 & 0 & 0 & \tilde{\beta} & 0 & 0 & 0 & 0 & 0 \\
    0 & 0 & 0 & 0 & 0 & 0 & 0 & \tilde{\beta} & 0 & 0 \\
    0 & 0 & 0 & 0 & 0 & 0 & -\tilde{\beta} & 0 & 0 & 0 \\
    0 & 0 & 0 & 0 & 0 & 0 & 0 & 0 & 0 & 0 \\
    0 & 0 & 0 & 0 & 0 & 0 & 0 & 0 & 0 & 0
  \end{pmatrix}
\end{equation}
It is natural to use the same coordinates introduced in
\eqref{eq:F5coords}, this time to parametrise the 4-plane spanned by
$x^1,\dots ,x^4$
\begin{equation}
%  \label{eq:F1acoords}
  x^1 + i x^2 = r \cos\theta e^{i(\psi + \varphi)} \qquad \text{and}
  \qquad x^3 + i x^4 = r \sin\theta e^{i(\psi - \varphi)}~,
\end{equation}
where $\theta \in [0,\frac\pi2]$, $\psi\in [0,\pi]$ and
$\varphi\in \RR/2\pi\ZZ$, and proceed analogously with the second
4-plane spanned by $x^5,\dots ,x^8$
\begin{equation}
%  \label{eq:F1acoords}
  x^5 + i x^6 = \tilde{r} \cos\tilde{\theta} 
e^{i(\tilde{\psi} + \tilde{\varphi})} \qquad \text{and}
  \qquad x^7 + i x^8 = \tilde{r} \sin\tilde{\theta} 
e^{i(\tilde{\psi} - \tilde{\varphi})}~,
\end{equation}
where $\tilde{\theta} \in [0,\frac\pi2]$, $\tilde{\psi}\in [0,\pi]$ and
$\tilde{\varphi}\in \RR/2\pi\ZZ$. The full distance to the fluxstring
is measured by the square root of $r^2 + \tilde{r}^2$. In terms of the 
new coordinates, the scalar function becomes
\begin{equation}
 \label{eq:F1alambda}
  \Lambda=1+r^2\beta^2+ \tilde{r}^2\tilde{\beta}^2
\end{equation}
fully determining the dilaton, $\phi=\frac{3}{4}\log\Lambda$, whereas 
the RR 1-form potential is given by
\begin{equation}
  \label{eq:F1aRR1form}
  A = \Lambda^{-1}\beta r^2 \left( d\varphi + \cos(2\theta)d\psi\right)
  + \Lambda^{-1}\tilde{\beta} \tilde{r}^2 \left( d\tilde{\varphi} 
  + \cos(2\tilde{\theta})d\tilde{\psi}\right)~,
\end{equation}
and the metric becomes
\begin{multline}
  \label{eq:F1ametric}
  g = \Lambda^{1/2} \biggl[ ds^2(\EE^{1,1}) + dr^2 + r^2 d\theta^2
  + r^2 \sin^2(2\theta)d\psi^2 \\
  + d\tilde{r}^2 + \tilde{r}^2 d\tilde{\theta}^2 +
  \tilde{r}^2 \sin^2(2\tilde{\theta})d\tilde{\psi}^2 \biggr]\\
  + \Lambda^{-1/2} \biggl[r^2 \left(d\varphi + \cos 2\theta
    d\psi\right)^2 (1+\tilde{\beta}\tilde{r}^2) + \tilde{r}^2
  \left(d\tilde{\varphi} + \cos 2\tilde{\theta} d\tilde{\psi}\right)^2
  (1+\beta r^2)\\
  -2\beta\tilde{\beta}r^2\tilde{r}^2\left(d\varphi + \cos 2\theta
    d\psi\right)\left(d\tilde{\varphi} + \cos 2\tilde{\theta}
    d\tilde{\psi} \right)\biggr]~.
\end{multline}

Finally, let us consider the case in which $B$ belongs to a Cartan
subalgebra of $\fsu(4)\subset \fso(8) \subset \fso(9,1)$:
\begin{equation}
%  \label{eq:F1bmatrix}
  B= 
  \begin{pmatrix}
    0 & -\beta_1 & 0 & 0 & 0 & 0 & 0 & 0 & 0 & 0 \\
    \beta_1 & 0 & 0 & 0 & 0 & 0 & 0 & 0 & 0 & 0 \\
    0 & 0 & 0 & -\beta_2 & 0 & 0 & 0 & 0 & 0 & 0 \\
    0 & 0 & \beta_2 & 0 & 0 & 0 & 0 & 0 & 0 & 0 \\
    0 & 0 & 0 & 0 & 0 & -\beta_3 & 0 & 0 & 0 & 0 \\
    0 & 0 & 0 & 0 & \beta_3 & 0 & 0 & 0 & 0 & 0 \\
    0 & 0 & 0 & 0 & 0 & 0 & 0 & -\beta_4 & 0 & 0 \\
    0 & 0 & 0 & 0 & 0 & 0 & \beta_4 & 0 & 0 & 0 \\
    0 & 0 & 0 & 0 & 0 & 0 & 0 & 0 & 0 & 0 \\
    0 & 0 & 0 & 0 & 0 & 0 & 0 & 0 & 0 & 0
  \end{pmatrix}
\end{equation}
The solution is again given explicitly by \eqref{eq:fluxbrane} with
\begin{equation}
  \Lambda = 1 + \sum_{i=1}^4\beta_i^2 \vert z_i \vert^2 \quad
  \text{and} \quad A = \frac{\Lambda^{-1}}{2i}
  \sum_{i=1}^4\beta_i\left(\bar{z}_idz_i - z_i d\bar{z}_i\right)~,
\end{equation}
where $z_i$ $i=1,2,3,4$ correspond to complex coordinates in the 8-plane
spanned by the real coordinates $x^1,\dots ,x^8$ transverse to the fluxbrane.
We can rewrite this in terms of the radial distance to the fluxbrane
by introducing coordinates:
\begin{equation}
% \label{eq:F1bcoords}
  \begin{aligned}[m]
    x^1+ix^2&=z_1=r\cos\theta_1 e^{i\varphi_1}\\
    x^3+ix^4&=z_2=r\sin\theta_1\cos\theta_2 e^{i\varphi_2} \\
    x^5+ix^6&=z_3=r\sin\theta_1\sin\theta_2\cos\theta_3
    e^{i\varphi_3}\\
    x^7+ix^8&=z_4=r\sin\theta_1 \sin\theta_2\sin\theta_3
    e^{i\varphi_4}~,
\end{aligned}
\end{equation}
where $\theta_i \in [0,\frac\pi2]$ and $\varphi_i \in \RR/2\pi\ZZ$ for
all $i=1,2,3,4$.  In terms of the new coordinates, the scalar function
becomes
\begin{multline}
 \label{eq:F1blambda}  
  \Lambda= 1 + r^2\left(\beta_1^2\cos^2\theta_1 +
    \beta_2^2\sin^2\theta_1 \cos^2\theta_2 \right.\\
  \left. + \beta_3^2\sin^2\theta_1\sin^2 \theta_2\cos^2\theta_3
    +\beta_4^2\sin^2\theta_1 \sin^2\theta_2\sin^2\theta_3\right)~.
\end{multline}
The latter fixes the dilaton to be $\phi=\frac{3}{4}\log\Lambda$, whereas 
the RR 1-form potential is given by
\begin{multline}
  \label{eq:F1bRR1form}
  A = \Lambda^{-1}r^2\left( \beta_1 \cos^2\theta_1 d\varphi_1 +
    \beta_2 \sin^2\theta_1\cos^2\theta_2 d\varphi_2\right.\\
  \left. +   \beta_3\sin^2\theta_1\sin^2 \theta_2\cos^2\theta_3
  d\varphi_3 + \beta_4\sin^2\theta_1 \sin^2\theta_2\sin^2\theta_3
  d\varphi_4\right)~,
\end{multline}
and the metric by
\begin{footnotesize}
\begin{multline}
  \label{eq:F1bmetric}
  g = \Lambda^{1/2} \left[ ds^2(\EE^{1,1}) + dr^2 +
    r^2\left(d\theta_1^2 +\sin^2\theta_1 d\theta_2^2 +
      \sin^2\theta_1\sin^2\theta_2 d\theta_3^2 \right) \right. \\
  + r^2 \left(\cos^2\theta_1 d\varphi_1^2 +
    \sin^2\theta_1\cos^2\theta_2 d\varphi_2^2 +
    \sin^2\theta_1\sin^2\theta_2\cos^2\theta_3 d\varphi_3^2 \right. \\
    \left. \left. + \sin^2\theta_1\sin^2\theta_2\sin^2\theta_3 d\varphi_4^2
    \right)\right]\\ 
  -\Lambda^{-1/2} r^4\left[ \beta_1 \cos^2\theta_1 d\varphi_1 +
    \beta_2 \sin^2\theta_1\cos^2\theta_2 d\varphi_2 +
    \beta_3\sin^2\theta_1\sin^2 \theta_2\cos^2\theta_3 d\varphi_3
  \right.\\
  \left. + \beta_4\sin^2\theta_1 \sin^2\theta_2\sin^2\theta_3
  d\varphi_4\right]^2~.
\end{multline}
\end{footnotesize}

\subsection{Supersymmetric nullbranes}

On the other extreme, we have these reductions where $u$ is different
from zero, but the $\beta$'s are zero.  These reductions give rise to
solutions with null RR field strengths which we tentatively call
\emph{nullbranes}.  In this case we have the freedom to choose a
coordinate system in which the null rotation is along one of the
coordinates, say $x^1$.  The corresponding matrix $B$ has all
$\beta$'s equal to zero and only $u$ is different from zero.
Moreover, it is possible to reabsorb $u$, provided it is nonzero, by
rescaling $x^\pm \mapsto u^{\pm 1} x^\pm$.  The resulting solution
seems to be new.\footnote{The possibility of quotienting by boosts was
  mentioned briefly in \cite{GSflux}.  The resulting fluxbranes are
  electrically charged but are not supersymmetric.}  Explicitly, we
have
\begin{equation}
  \label{eq:N7LA}
  \Lambda = 1 + (x^-)^2 \quad\text{and}\quad
  A = \frac{1}{1 + (x^-)^2} \left(x^- dx^1 - x^1 dx^-\right)~,
\end{equation}
a non-trivial dilaton $\phi=\frac{3}{4}\log\Lambda$ and a type IIA metric 
in the string frame given by
\begin{multline}
  \label{eq:N7metric}
  g = \Lambda^{1/2} \left[2 dx^+ dx^- - (x^1)^2 (dx^-)^2 +
    ds^2\left(\EE^{7}\right)\right]\\
  + \Lambda^{-1/2} \left(dx^1 + x^1 x^- dx^-\right)^2~.
\end{multline}
The RR $2$-form field strength is null
\begin{equation}
  \label{eq:N7nullF}
  F = \frac{2}{\Lambda^2} dx^- \wedge dx^1~,
\end{equation}
hence our name for these solutions.  These solutions always preserve
one half of the supersymmetry.

Notice that the metric \eqref{eq:N7metric} has an $\EE^7$ subspace, so
this suggests computing the Hodge dual of the null
field strength \eqref{eq:N7nullF}.  Indeed, $\star F$ couples
naturally to a seven-dimensional extended object.  Doing so one finds
that $\star F$ is actually constant in this coordinate system and is
given by
\begin{equation}
  \star F = 2 dx^- \wedge \dvol(\EE^7)
\end{equation}
for our choice of orientation.  This is reminiscent of the homogeneous
branes (H-branes) introduced in \cite{FOPflux} (see also
\cite{NewIIB}).

\subsection{Generalised supersymmetric fluxbranes}

When the matrix $B$ contains both rotations and null rotations, the
resulting solution interpolates between the fluxbranes and the
nullbranes discussed above.  To illustrate the geometry of these
solutions, let us consider a matrix $B$ which is the sum of the matrix
\eqref{eq:F5matrix} giving rise to the F5-brane and the matrix giving
rise to the N7-brane:
\begin{equation}
  B= 
  \begin{pmatrix}
    0 & 0 & 0 & 0 & 0 & 0 & 0 & 0 & 0 & u \\
    0 & 0 & 0 & 0 & 0 & 0 & 0 & 0 & 0 & 0 \\
    0 & 0 & 0 & 0 & 0 & 0 & 0 & 0 & 0 & 0 \\
    0 & 0 & 0 & 0 & 0 & 0 & 0 & 0 & 0 & 0 \\
    0 & 0 & 0 & 0 & 0 & -\beta & 0 & 0 & 0 & 0 \\
    0 & 0 & 0 & 0 & \beta & 0 & 0 & 0 & 0 & 0 \\
    0 & 0 & 0 & 0 & 0 & 0 & 0 & \beta & 0 & 0 \\
    0 & 0 & 0 & 0 & 0 & 0 & -\beta & 0 & 0 & 0 \\
    -u & 0 & 0 & 0 & 0 & 0 & 0 & 0 & 0 & 0 \\
    0 & 0 & 0 & 0 & 0 & 0 & 0 & 0 & 0 & 0
  \end{pmatrix}~.
\end{equation}
As in the case of the nullbrane, one can reabsorb $u$
by a rescaling $x^\pm \mapsto u^{\pm 1} x^\pm$.  Applying the general
formulae \eqref{eq:fluxLA} and \eqref{eq:fluxbrane}, it is easy to
write down the resulting IIA background.  In the coordinates
\eqref{eq:F5coords}, the dilaton is again given by $\phi = \frac34
\log\Lambda$, the RR 1-form potential by
\begin{equation}
   A = \Lambda^{-1} \left( x^-dx^1 - x^1 dx^- + \beta
   r^2 \left(d\varphi + \cos 2\theta d\psi\right)\right)
\end{equation}
and the metric by
\begin{footnotesize}
  \begin{multline}
    g = \Lambda^{1/2}\left[2dx^+ dx^- - (x^1)^2(dx^-)^2 + ds^2(\EE^3)
      + dr^2 + r^2(d\theta^2 + \sin^2 2\theta d\psi^2)\right]\\
    + \Lambda^{-1/2}\biggl[\left(dx^1 + x^1 x^- dx^-\right)^2 +
    r^2\left(d\varphi + \cos 2\theta d\psi\right)^2
    \left(1+(x^-)^2\right) \\
    + (\beta r)^2\left((dx^1)^2 + (x^1)^2(dx^-)^2\right) -2 \beta r^2
    (x^-dx^1 - x^1 dx^-)(d\varphi + \cos 2\theta d\psi)\biggr]~,
  \end{multline}
\end{footnotesize}
where the scalar function is $\Lambda= 1 + (x^-)^2 + \beta^2 r^2$.

This solution smoothly interpolates between the F5-brane---which is
recovered in the region $x^- \ll 1$ keeping $\beta r$ fixed---and
the N7-brane, in the region $\beta r\ll 1$ keeping $x^-$ fixed.  It
preserves $1/4$ of the spacetime supersymmetry. 

There is one more class of solutions which interpolates between the
N7-brane and the F3-brane.  In this case the matrix B is the sum of
the matrix \eqref{eq:F3matrix} and the matrix giving rise to the N7-brane
\begin{equation}
  B= 
  \begin{pmatrix}
    0 & 0 & 0 & 0 & 0 & 0 & 0 & 0 & 0 & u \\
    0 & 0 & 0 & 0 & 0 & 0 & 0 & 0 & 0 & 0 \\
    0 & 0 & 0 & -\beta_2 & 0 & 0 & 0 & 0 & 0 & 0 \\
    0 & 0 & \beta_2 & 0 & 0 & 0 & 0 & 0 & 0 & 0 \\
    0 & 0 & 0 & 0 & 0 & -\beta_3 & 0 & 0 & 0 & 0 \\
    0 & 0 & 0 & 0 & \beta_3 & 0 & 0 & 0 & 0 & 0 \\
    0 & 0 & 0 & 0 & 0 & 0 & 0 & \beta_4 & 0 & 0 \\
    0 & 0 & 0 & 0 & 0 & 0 & -\beta_4 & 0 & 0 & 0 \\
    -u & 0 & 0 & 0 & 0 & 0 & 0 & 0 & 0 & 0 \\
    0 & 0 & 0 & 0 & 0 & 0 & 0 & 0 & 0 & 0
  \end{pmatrix}~.
\end{equation}
The solution has three parameters since $\beta_2+\beta_3+\beta_4=0$, 
but as argued before, one can reabsorb $u$ by a rescaling 
$x^\pm \mapsto u^{\pm 1} x^\pm$. The type IIA configuration can be written
down by proceeding as before; applying the general formulae \eqref{eq:fluxLA} 
and \eqref{eq:fluxbrane}, and this time using the coordinates 
\eqref{eq:F3coords}. The dilaton is given by 
$\phi = \frac{3}{4}\log\Lambda$, the RR 1-form potential by
\begin{multline}
   A = \Lambda^{-1} \left( x^-dx^1 - x^1 dx^- + r^2 
   \left(\beta_2 \cos^2\theta_1 d\varphi_1 \right.\right. \\
   \left.\left. +\beta_3 \sin^2\theta_1\cos^2\theta_2 d\varphi_2 + 
   \beta_4\sin^2\theta_1\sin^2 \theta_2 d\varphi_3\ \right)\right)
\end{multline}
and the metric by
\begin{footnotesize}
  \begin{multline}
     g = \Lambda^{1/2}\left[2dx^+ dx^- + (dx^1)^2 + (dx^2)^2 +
     + dr^2 + r^2 \left(d\theta_1^2 + \sin^2 \theta_1 d\theta_2^2\right) 
     \right. \\
     \left. + r^2\left(\cos^2\theta_1 d\varphi_1^2 +
     \sin^2\theta_1\cos^2\theta_2 d\varphi_2^2 +
     \sin^2\theta_1\sin^2\theta_2 d\varphi_3^2\right) \right] \\
     -\Lambda^{-1/2} \left[r^4 \left(\beta_2 \cos^2\theta_1 d\varphi_1 +
    \beta_3 \sin^2\theta_1\cos^2\theta_2 d\varphi_2 +
    \beta_4\sin^2\theta_1\sin^2 \theta_2 d\varphi_3\right)^2\right. \\    
    \left. +\left(x^-dx^1-x^1dx^-\right)^2 + 2r^2 \left(x^-dx^1-x^1dx^-\right)
    \cdot \right. \\
    \left. \cdot  \left(\beta_2 \cos^2\theta_1 d\varphi_1 +
    \beta_3 \sin^2\theta_1\cos^2\theta_2 d\varphi_2 +
    \beta_4\sin^2\theta_1\sin^2 \theta_2 d\varphi_3\right)\right]~.,
  \end{multline}
\end{footnotesize}
where the scalar function is 
\begin{footnotesize}
  \begin{equation*}
     \Lambda = 1 + (x^-)^2 + r^2\left(\beta_2^2\cos^2\theta_1 +
     \beta_3^2\sin^2\theta_1 \cos^2\theta_2 +
     \beta_4^2\sin^2\theta_1\sin^2 \theta_2\right)~.
  \end{equation*}
\end{footnotesize} 
The above configuration preserves $\tfrac18$ of the supersymmetry of
the eleven-dimensional vacuum.  It smoothly interpolates between the
F3-brane ---which is recovered in the region $x^-\ll 1$ keeping
$\beta_i\,r$ fixed for all $i$--- and the N7-brane, in the region
$\beta_i\,r\ll 1$ $\forall\,i$ keeping $x^-$ fixed. Notice that in
both interpolating solutions presented in this subsection, as we move
in the parameter space we find regions of supersymmetry enhancement,
corresponding to the N7-brane and the F5/F3-branes themselves.

\section{Dualities and flux branes}
\label{sec:dualities}

Any supergravity background gives rise to a family of related
solutions through the use of U-duality transformations and other
solution-generating techniques.  The purpose of this section is to
discuss the family of solutions obtained by U-duality from some of the
generalised fluxbranes found in the previous section, namely
fluxbranes and nullbranes.  It is well known that any time a T-duality
transformation is used to generate a new solution, the latter is
delocalised in the T-dual direction.  It is often possible to find
localised versions of the new solutions, but it remains to be seen
whether this is indeed true for the solutions we find here.  We will
make much use of the T-duality rules derived by Bergshoeff, Hull and
Ortín in \cite{BHO}.

\subsection{A word on the notation}

In applying the T-duality rules to the solutions found above, we will
obtain many solutions for which no name yet exists.  In this section
we have introduced a notation in order to be able to identify them.
Since the notation is not standard, let us take a moment to explain
it.  By an $\text{F}(p,q)$-brane, we shall denote a solution with full
Poincaré invariance in $p+1$ dimensions, but which is nevertheless
delocalised in $q$ of them.  A similar notation, now $N(p,q)$ is
employed in the section on nullbrane dualities.  Whenever an S-duality
transformation is applied, an $s$ is added to the notation together
with one of the letters $a$ or $b$ to emphasise that this is a
configuration of type IIA or IIB, respectively.  This is done to
distinguish solutions which, although formally equivalent, belong to
different theories.  In the section on fluxstring dualities, we
further adorn the notation with a subscript $1$ or $2$, to distinguish
between the two fluxstring configurations.  The rest of the notation
is standard.

\subsection{F5-brane dualities}

The starting solution is the F5-brane, with metric \eqref{eq:F5metric}
and RR 1-form potential \eqref{eq:F5RR1form}, whereas the dilaton is
given by $\phi = \frac34 \log (1 + \beta^2 r^2)$.

By performing a T-duality along $5-p$ worldspace directions of the
F5-brane, one obtains a family of $\text{F}(p,5-p)$-brane
configurations, for $p=4,3,2,1,0$, belonging to type IIA for odd $p$
and to type IIB for even $p$.  They can be jointly described by
\begin{equation}
  \label{p5-p}
  \begin{aligned}[m]
    \phi &= \frac{p-2}{4}\log\Lambda~, \qquad G_{(7-p)}= F \wedge
    \dvol(\EE^{5-p}) \\
    g &= \Lambda^{1/2}ds^2(\EE^{p,1}) + \Lambda^{-1/2}ds^2(\EE^{5-p})
    + \Lambda^{1/2}h~,
  \end{aligned}
\end{equation}  
where the transverse metric $h$ is defined as
\begin{equation}
  h = dr^2 + r^2(d\theta^2 + \sin^2 2\theta d\psi^2)+
  \frac{r^2}{\Lambda}(d\varphi + \cos 2\theta d\psi)^2~,
\end{equation}
and $G_{(7-p)}$ stands for the RR field strength, $F$ being the RR 2-form
field strength of the original F5-brane configuration.

Notice that the $\text{F}(2,3)$-brane is S-selfdual, whereas the
$\text{F}(4,1)$-brane is S-dual to the $\text{F}(4,1)bs$-brane
\begin{equation}
  \label{41bs}
  \begin{aligned}[m]
    \phi &= -\half\log\Lambda~, \qquad H=-F\wedge dx\\
    g &= ds^2(\EE^{4,1}) + \Lambda^{-1}dx^2 + h~,
  \end{aligned}
\end{equation}
where $H$ is the NS-NS three-form field strength.  Similarly, the
$\text{F}(0,5)$-brane is S-dual to the $\text{F}(0,5)bs$-brane
\begin{equation}
  \label{05bs}
  \begin{aligned}[m]
    \phi &= \half\log\Lambda~, \qquad \star H= F\wedge
    \dvol(\EE^5) \\
    g &= \Lambda [-dt^2 + h] + ds^2(\EE^5)~.
  \end{aligned}
\end{equation}

By a T-duality along $\RR^4$ in the $\text{F}(4,1)bs$-brane, we obtain
the $\text{F}(4,1)$ as-brane, which is formally equivalent to its
T-dual.  Both are T-duals along the twisted directions to the
corresponding vacuum solutions in type IIA/B with non-trivial
identifications.  Similarly, the new $\text{F}(0,5)bs$ -brane
configuration also generates through T-duality a new type IIA
solution, the $\text{F}(0,5)as$-brane, which is again equivalent to
its type IIB T-dual.

\begin{figure}
  \begin{center}
    \includegraphics[angle=90]{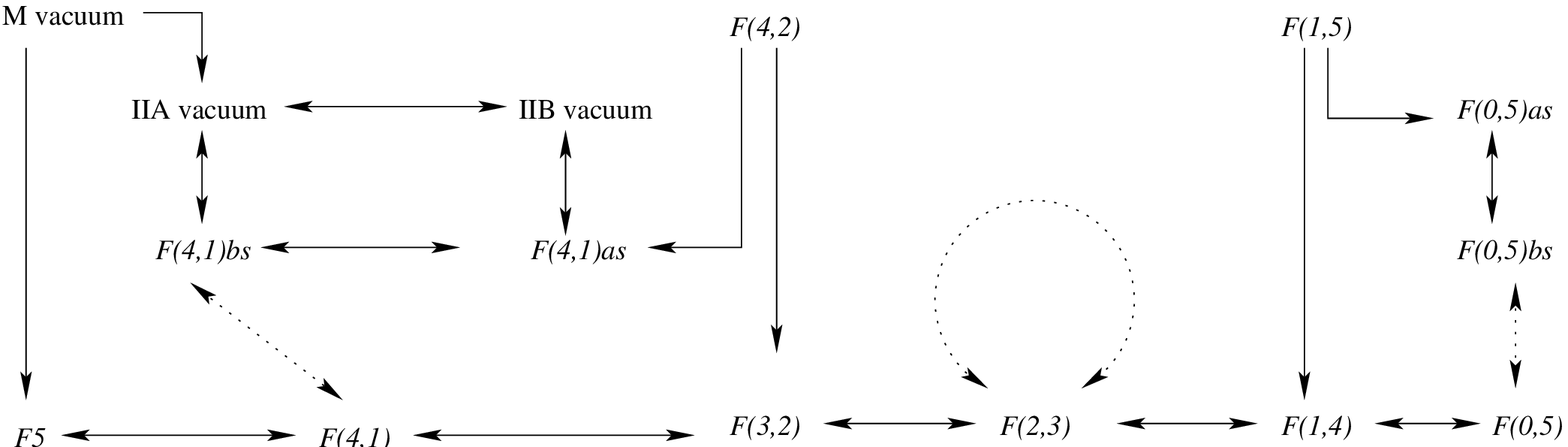}
    \caption{Web of dualities associated with an F5-brane.
      A dotted line represents an S-duality transformation, a
      unidirectional solid line represents a Kaluza--Klein reduction
      from M-theory, and a bidirectional solid line a T-duality
      transformation.}
    \label{fig:F5fig}
  \end{center}
\end{figure}

Once new type IIA configurations have been generated, they can also be
understood as the Kaluza--Klein reductions of certain new M-theory
configurations.  In particular, the $\text{F}(4,1)as$-brane and the
$\text{F}(3,2)$-brane are the reductions of the
$\text{F}(4,2)$-brane
\begin{equation}
  \label{m42}
  \begin{aligned}[m]
    F_{(4)} &= F\wedge \dvol(\EE^2)\\
    ds_{11}^2 &= \Lambda^{1/3}[ds^2(\EE^{4,1})+h] +
    \Lambda^{-2/3}ds^2(\EE^2)~,
  \end{aligned}
\end{equation}
whereas the $\text{F}(1,4)$-brane and the $\text{F}(0,5)as$ -brane come
from the $\text{F}(1,5)$-brane
\begin{equation}
  \label{m15}
  \begin{aligned}[m]
    \star F_{(4)} &= F \wedge \dvol(\EE^5) \\
    ds^2_{11} &= \Lambda^{2/3} [-ds^2(\EE^{1,1}) + h] +
    \Lambda^{-1/3}ds^2(\EE^5)~.
  \end{aligned}
\end{equation}
This is illustrated in Figure~\ref{fig:F5fig}.

\subsection{F3-brane dualities}

The starting point is the F3-brane, with metric \eqref{eq:F3metric}
and RR 1-form potential \eqref{eq:F3RR1form}, whereas the dilaton is
given by $\phi =\frac{3}{4}\log\Lambda$, $\Lambda$ being the one
defined in \eqref{eq:F3lambda}.

Proceeding as in the previous subsection, one can generate a family of
$\text{F}(p,3-p)$-brane configurations by performing T-duality
transformations along $3-p$ of the the worldspace directions of the
original F3-brane. They belong to type IIA for odd $p$ and to type IIB
for even $p$. They are jointly described by
\begin{equation}
  \label{p3pbrane}
  \begin{aligned}[m]
    \phi &= \frac{p}{4}\log\Lambda~, \qquad G_{(5-p)} = F\wedge
    \dvol(\EE^{3-p})\\
    g &= \Lambda^{1/2}ds^2(\EE^{p,1}) +\Lambda^{-1/2}ds^2(\EE^{3-p}) +
    h ~,
  \end{aligned}
\end{equation}
where the transverse metric $h$ is defined by
\begin{footnotesize}
  \begin{multline}
    h= \Lambda^{1/2} \left[ dr^2 + r^2
      \left(d\theta_1^2 + \sin^2 \theta_1 d\theta_2^2\right) \right. \\
    \left. + r^2\left(\cos^2\theta_1 d\varphi_1^2 +
        \sin^2\theta_1\cos^2\theta_2 d\varphi_2^2 +
        \sin^2\theta_1\sin^2\theta_2 d\varphi_3^2\right) \right]\\
    - \Lambda^{-1/2} r^4 \left(\beta_2 \cos^2\theta_1 d\varphi_1 +
      \beta_3 \sin^2\theta_1\cos^2\theta_2 d\varphi_2 +
      \beta_4\sin^2\theta_1\sin^2 \theta_2 d\varphi_3\right)^2~.
  \end{multline}
\end{footnotesize}

Notice that the $\text{F}(0,3)$-brane is S-selfdual---in particular,
it has a constant dilaton---whereas the $\text{F}(2,1)$-brane is
S-dual to the $\text{F}(2,1)bs$-brane described by
\begin{equation}
  \begin{aligned}[m]
    \phi &= -\half\log\Lambda~, \qquad H=-F\wedge dx \\
    g &= ds^2(\EE^{2,1}) + \Lambda^{-1} dx^2 + \Lambda^{-1/2} h~.
  \end{aligned}
\end{equation}
Using T-duality along the longitudinal $\EE^2$ subspace, one generates
the same solution in type IIA, namely the $\text{F}(2,1)as$-brane.
Both are the T-duals of the corresponding IIA/B vacuum when applying a
T-duality transformation along the twisted direction.

\begin{figure}
  \begin{center}
    \includegraphics[angle=90]{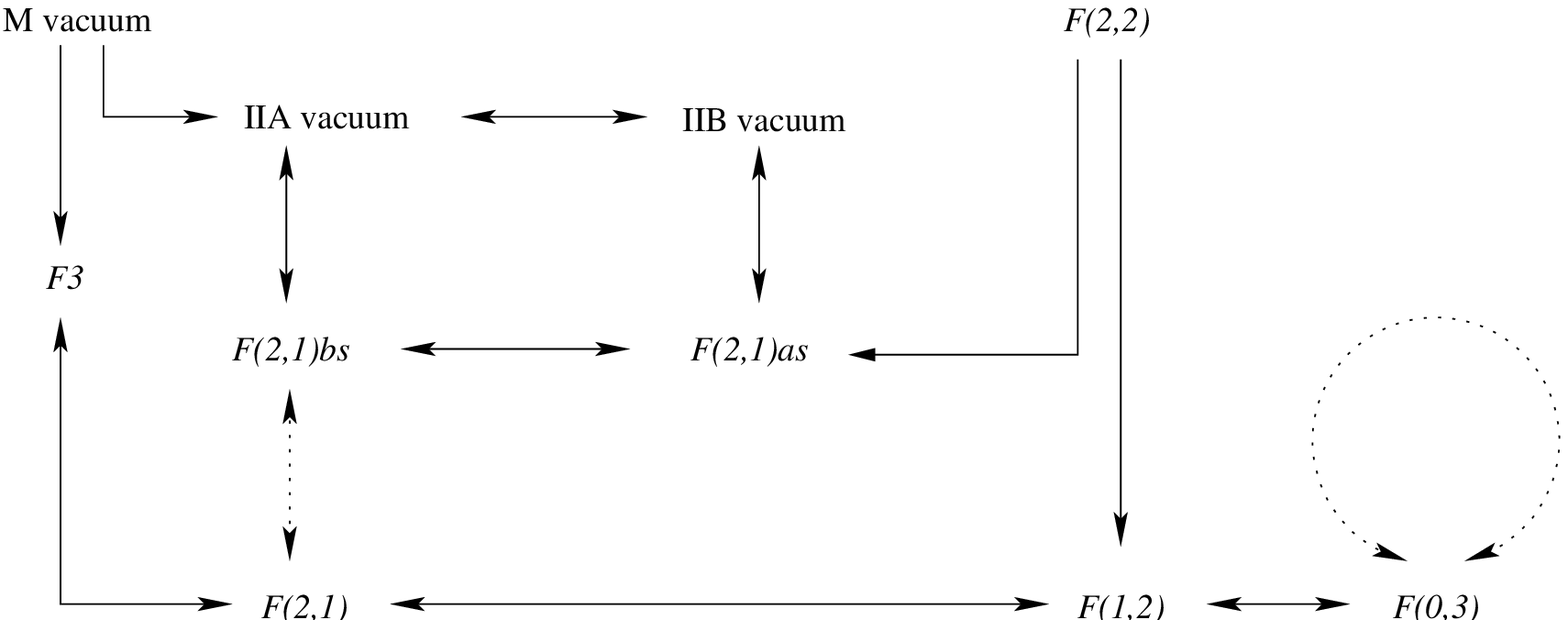}
    \caption{Web of dualities associated with an F3-brane.
      A dotted line represents an S-duality transformation, a
      unidirectional solid line represents a Kaluza--Klein reduction
      from M-theory, and a bidirectional solid line a T-duality
      transformation.}
    \label{fig:F3fig}
  \end{center}
\end{figure}

Finally, both the $\text{F}(1,2)$-brane and the
$\text{F}(2,1)as$-brane can be understood as the Kaluza--Klein
reduction of the $\text{F}(2,2)$-brane in eleven dimensions
\begin{equation}
  \label{22mbrane}
  \begin{aligned}[m]
    F_{(4)} &= F\wedge \dvol(\EE^2)  \\
    ds_{(11)}^2 &= \Lambda^{1/3}ds^2(\EE^{2,1}) +
    \Lambda^{-2/3}ds^2(\EE^2) + \Lambda^{-1/6} h~.
  \end{aligned}
\end{equation}
This finishes the web of dualities for the F3-brane, which is
summarised in Figure~\ref{fig:F3fig}.

\subsection{F1-string dualities}

Due to the fact that there exist two inequivalent fluxstring
configurations preserving different amount of supersymmetry, we shall
introduce some notation that will allow us to discuss their U-duality
properties in a joint discussion. In particular, we shall introduce an
index $i=1,2$ such that all quantities indexed by $i=1$ will
correspond to the fluxstring preserving $1/4$ supersymmetry, whereas
those indexed by $i=2$ will describe the $1/8$ configuration.

Performing a T-duality transformation along its worldspace direction,
one generates the type IIB $\text{F}(0,1)_i$-string
\begin{equation}
  \label{01string}
  \begin{aligned}[m]
    \phi_i &= \half\log\Lambda_i~, \qquad G^i_{(3)} = F^i\wedge dx \\
    g_i &= -\Lambda_i^{1/2}dt^2 + \Lambda_i^{-1/2} dx^2 + h_i~,
  \end{aligned}
\end{equation}
where $\Lambda_1$ is defined in \eqref{eq:F1alambda}, $F^1$ is the RR
2-form field strength derived from \eqref{eq:F1aRR1form} and $h_1$ is
the transverse metric
\begin{footnotesize}
  \begin{multline}
    h_1 = \Lambda_1^{1/2} \biggl[ dr^2 + r^2 d\theta^2
      + r^2 \sin^2(2\theta)d\psi^2 \\
    + d\tilde{r}^2 + \tilde{r}^2 d\tilde{\theta}^2 +
      \tilde{r}^2 \sin^2(2\tilde{\theta})d\tilde{\psi}^2 \biggr]\\
    + \Lambda_1^{-1/2} \biggl[r^2 \left(d\varphi + \cos 2\theta
        d\psi\right)^2 (1+\tilde{\beta}\tilde{r}^2) + \tilde{r}^2
      \left(d\tilde{\varphi} + \cos 2\tilde{\theta}
        d\tilde{\psi}\right)^2
      (1+\beta r^2)\\
    -2\beta\tilde{\beta}r^2\tilde{r}^2\left(d\varphi + \cos
        2\theta d\psi\right)\left(d\tilde{\varphi} + \cos
        2\tilde{\theta} d\tilde{\psi} \right)\biggr]~.
  \end{multline}
\end{footnotesize}
Similarly, $\Lambda_2$ is defined in \eqref{eq:F1blambda}, $F^2$ is
the RR 2-form field strength derived from \eqref{eq:F1bRR1form} and
$h_2$ is the transverse metric
\begin{footnotesize}
  \begin{multline}
    h_2 = \Lambda_2^{1/2} \left[ dr^2 +
      r^2\left(d\theta_1^2 +\sin^2\theta_1 d\theta_2^2 +
        \sin^2\theta_1\sin^2\theta_2 d\theta_3^2 \right) \right. \\
    + r^2 \left(\cos^2\theta_1 d\varphi_1^2 +
      \sin^2\theta_1\cos^2\theta_2 d\varphi_2^2 +
      \sin^2\theta_1\sin^2\theta_2\cos^2\theta_3 d\varphi_3^2 \right. \\
    \left. \left. + \sin^2\theta_1\sin^2\theta_2\sin^2\theta_3 d\varphi_4^2
      \right)\right]\\ 
    -\Lambda_2^{-1/2} r^4\left[ \beta_1 \cos^2\theta_1 d\varphi_1 +
      \beta_2 \sin^2\theta_1\cos^2\theta_2 d\varphi_2 +
      \beta_3\sin^2\theta_1\sin^2 \theta_2\cos^2\theta_3 d\varphi_3
    \right.\\
    \left. + \beta_4\sin^2\theta_1 \sin^2\theta_2\sin^2\theta_3
      d\varphi_4\right]^2~.
  \end{multline}
\end{footnotesize}
Close to the core, that is, for $r^2\beta_i^2 \ll 1$ for all $i$,
where the supergravity description is reliable since the string
coupling constant is weak, \eqref{01string} is flat, whereas far away
from the core, that is, for $r^2\beta_i^2 \gg 1$ for all $i$, it is no
longer conformal to a cylindrical metric. It is, in any case, not
appropriate to use the supergravity description in that region since
the string coupling constant blows up there and one should take into
account higher order corrections in $g_s$ into the classical
supergravity equations of motion or use its S-dual description, which
is the $\text{F}(0,1)_ibs$-string:
\begin{equation}
  \begin{aligned}[m]
    \label{01bsstring}
    \phi_i &= -\half\log\Lambda_i~, \qquad H_i= -F^i \wedge dx \\
    g &= -dt^2 + \Lambda_i^{-1} dx^2 +\Lambda_i^{-1/2}h_i~.
  \end{aligned}
\end{equation} 
Notice that the above solution is equivalent to the type IIB configuration 
obtained by applying a T-duality transformation along the twisted direction 
on the type IIA vacuum with topological non-trivial identifications.

\begin{figure}
  \begin{center}
    \includegraphics{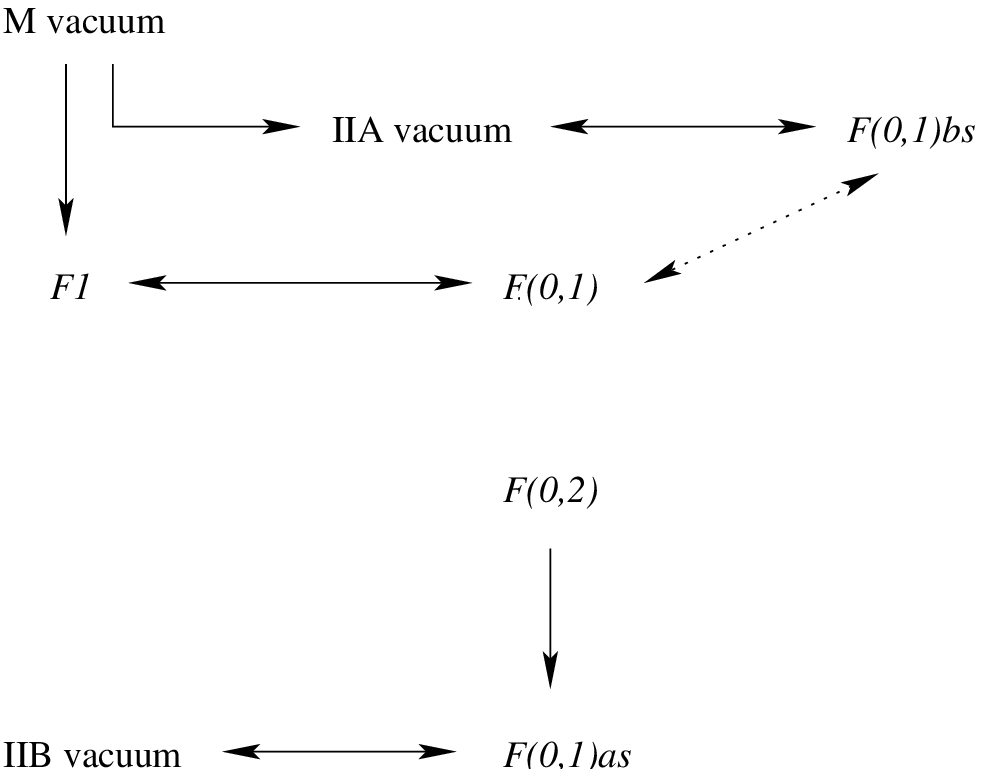}
    \caption{Web of dualities associated with a F1-string.
      A bold line represents an S-duality transformation, while
      unidirectional lines indicate Kaluza--Klein reductions from
      M-theory, and bidirectional lines a T-duality
      transformation.  This diagram is valid for both types of
      fluxstrings, adding the relevant subscript ($1$ or $2$, but
      always the same) to the $F(0,1)$ solutions.}
    \label{fig:F1fig}
  \end{center}
\end{figure}

Of course, one could have started from the flat background in type IIB
with non-trivial identifications and generate the
$\text{F}(0,1)_ias$-string , which is formally equivalent to
\eqref{01bsstring}. This new type IIA solution can be seen as the
Kaluza--Klein reduction of the $\text{F}(0,2)$-brane
\begin{equation}
  \label{02mbrane}
  \begin{aligned}[m]
    F^i_{(4)} &= -F^i \wedge \dvol(\EE^2) \\
    ds_{11}^2 &= -\Lambda_i^{1/3} dt^2 + \Lambda_i^{-2/3}ds^2(\EE^2)
    + \Lambda_i^{1/6} h_i ~.
  \end{aligned}
\end{equation} 
This finishes the web of dualities obtained from the F1-string
and which is summarised in Figure~\ref{fig:F1fig}.

\subsection{N7-brane dualities}

The purpose of this subsection is to give a preliminary analysis of
the family of solutions related to the N7-brane \eqref{eq:N7nullF}.
It will be convenient to introduce the following notation
\begin{equation}
  h = \Lambda^{1/2}\left(2dx^+ dx^- - (x^1)^2 (dx^-)^2\right)
  + \Lambda^{-1/2}\left(dx^1 + x^1 x^- dx^-\right)^2 ~.
\end{equation}
where $\Lambda$ is the one appearing in \eqref{eq:N7LA}.

By T-duality along $7-p$ of the spacelike directions in $\EE^7$ in
\eqref{eq:N7nullF}, one generates a family of N$(p,7-p)$-branes with
dilaton
\begin{equation}
  \phi = \frac{p-4}{4}\log \Lambda ~,
\end{equation}
RR field strength
\begin{equation}
  F_{(9-p)} = F \wedge \dvol(\EE^{7-p})~,
\end{equation}
and metric
\begin{equation}
  \label{eq:Np7-p}
  g = \Lambda^{1/2}ds^2(\EE^p) + \Lambda^{-1/2}ds^2(\EE^{7-p})
  + h ~,
\end{equation}
where $F$ is the one defined in \eqref{eq:N7nullF}.

One can then act with S-duality on the new type IIB configurations 
$(p=6,4,2,0)$. The N(4,3)-brane is S-selfdual (hence it has constant
dilaton) whereas the N(6,1)-brane is S-dual to the N$(6,1)bs$-brane
\begin{equation}
  \phi = -\half\log \Lambda \qquad \text{and} \qquad
  H= -F\wedge dx^7 ~,
\end{equation}
with metric
\begin{equation}
 \label{eq:N61bs}
 g = ds^2(\EE^6) + \Lambda^{-1}d(y^7)^2 + \Lambda^{-1/2}h ~.
\end{equation}
Similarly, the N(2,5)-brane is S-dual to the N$(2,5)bs$-brane
\begin{equation}
 \phi = \half\log \Lambda \qquad \text{and} \qquad
 H= 2 dx^-\wedge \dvol(\EE^2)~,
\end{equation}
with metric
\begin{equation}
 \label{eq:N25bs}
 g = \Lambda ds^2(\EE^2) + ds^2(\EE^5) + \Lambda^{1/2}h ~,
\end{equation}
whereas the N(0,7)-brane is S-dual to the N$(0,7)bs$-brane
\begin{equation}
 \phi = \log \Lambda \qquad \text{and} \qquad F_{(1)}= 2dx^-~,
\end{equation}
with metric
\begin{equation}
 \label{eq:N07bs}
 g = \Lambda^{1/2}ds^2(\EE^7) + \Lambda h ~.
\end{equation}

One can proceed systematically, uplifting the new type IIA
configurations obtained on the process to get new M-theory
configurations such as the N(6,2)-brane described by
\begin{equation}
 F_{(4)}= F\wedge \dvol(\EE^2)~,
\end{equation}
and metric
\begin{equation}
 \label{eq:N62bs}
 g = \Lambda^{1/3}ds^2(\EE^6) + \Lambda^{-2/3} ds^2(\EE^2) 
 + \Lambda^{-1/6} h~.
\end{equation}

There are many more configurations that go beyond the scope of our
present analysis, but deserve further study since, among many other
interests, they might lead to more general ansätze to solve the
eleven- and ten-dimensional supergravity equations of motion.

\section*{Acknowledgements}

Much of this work was done while the authors were participating in the
programme \emph{Mathematical Aspects of String Theory} at the Erwin
Schrödinger Institute in Vienna, whom we would like to thank for
support and for providing such a stimulating environment in which to
do research.  JMF is a member of EDGE, Research Training Network
HPRN-CT-2000-00101, supported by The European Human Potential
Programme.  JS is supported by a Marie Curie Fellowship of the
European Community programme ``Improving the Human Research Potential
and the Socio-Economic Knowledge Base'' under the contract number
HPMF-CT-2000-00480, and in part by a grant from the United
States--Israel Binational Science Foundation (BSF), the European RTN
network HPRN-CT-2000-00122 and by Minerva.  In addition, JMF would
like to acknowledge a travel grant from PPARC.

\bibliographystyle{amsplain}
\bibliography{AdS,ESYM,Sugra,Geometry,CaliGeo}

\end{document}